\documentclass[pra,twocolumn,preprintnumbers,superscriptaddress]{revtex4}
\usepackage{amssymb}
\usepackage{amsmath}
\usepackage{amsfonts}
\usepackage{wasysym}
\usepackage{graphicx}
\usepackage{dcolumn}
\usepackage{bm}
\usepackage{color}
\usepackage{ulem}
\usepackage{bbold}

\def\be{\begin{equation}}
\def\ee{\end{equation}}
\def\bea{\begin{eqnarray}}
\def\eea{\end{eqnarray}}

\begin{document}
\title{Unveiling quantum phases in quasi-one-dimensional dipolar gases using continuous matrix product state}
\author{Li Peng}
\affiliation{CAS Key Laboratory of Theoretical Physics, Institute of Theoretical Physics, Chinese Academy of Sciences, Beijing 100190, China}
\author{Junqiao Pan}
\affiliation{CAS Key Laboratory of Theoretical Physics, Institute of Theoretical Physics, Chinese Academy of Sciences, Beijing 100190, China}

\author{Su Yi}
\email{syi@itp.ac.cn}
\affiliation{CAS Key Laboratory of Theoretical Physics, Institute of Theoretical Physics, Chinese Academy of Sciences, Beijing 100190, China}
\affiliation{CAS Center for Excellence in Topological Quantum Computation \& School of Physical Sciences, University of Chinese Academy of Sciences, Beijing 100049, China}
\affiliation{Peng Huanwu Collaborative Center for Research and Education, Beihang University, Beijing 100191, China}

\author{Tao Shi}
\email{tshi@itp.ac.cn}
\affiliation{CAS Key Laboratory of Theoretical Physics, Institute of Theoretical Physics, Chinese Academy of Sciences, Beijing 100190, China}
\affiliation{CAS Center for Excellence in Topological Quantum Computation \& School of Physical Sciences, University of Chinese Academy of Sciences, Beijing 100049, China}
\affiliation{Peng Huanwu Collaborative Center for Research and Education, Beihang University, Beijing 100191, China}

\begin{abstract}
We investigate the ground-state properties of the quasi-one-dimensional dipolar gases using continuous matrix product states techniques. Making use of the first- and second-order correlation functions, we find that the system supports the superfluid, super-Tonks-Girardeau, and quasicrystal phases according to the Luttinger liquid theory. We also map out the phase diagram on the parameter plane consisting the contact and dipolar interaction strengths. Furthermore, we compute the Luttinger parameter, the structure factor, and the momentum distribution of the system. Finally, we show that the predicted dipolar effect can potentially be observed in quasi-one-dimensional gases of polar molecules.
\end{abstract}

\date{\today }
\maketitle

\affiliation{CAS Key Laboratory of Theoretical Physics, Institute of
Theoretical Physics, Chinese Academy of Sciences, Beijing 100190, China}

\affiliation{CAS Key Laboratory of Theoretical Physics, Institute of
Theoretical Physics, Chinese Academy of Sciences, Beijing 100190, China}

\affiliation{CAS Key Laboratory of Theoretical Physics, Institute of
Theoretical Physics, Chinese Academy of Sciences, Beijing 100190, China}
\affiliation{CAS Center for Excellence in Topological Quantum Computation \&
School of Physical Sciences, University of Chinese Academy of Sciences,
Beijing 100049, China}
\affiliation{Peng Huanwu Collaborative Center for
Research and Education, Beihang University, Beijing 100191, China}

\affiliation{CAS Key Laboratory of Theoretical Physics, Institute of
Theoretical Physics, Chinese Academy of Sciences, Beijing 100190, China}
\affiliation{CAS Center for Excellence in Topological Quantum Computation \&
School of Physical Sciences, University of Chinese Academy of Sciences,
Beijing 100049, China}
\affiliation{Peng Huanwu Collaborative Center for
Research and Education, Beihang University, Beijing 100191, China}

\section{Introduction}
Over the past two decades, significant developments have been achieved in experiments involving one-dimensional (1D) cold atomic gases~\cite{kinoshita2004, paredes2004,kinoshita2006,hofferberth2007,palzer2009, fabbri2012,kuhnert2013,ronzheimer2013, cazalilla2011,guan2013}. By trapping and cooling ultracold atomic gases, made it possible to  realized both bosonic and fermionic integrable model in experiment, such as Lieb-Liniger (LL) model~\cite{lieb1963} and Yang-Gaudin model~\cite{yang1967,gaudin1967}.
The experimental realization of the 1D Tonks-Girardeau (TG) gases and BCS pairs allow the studies of the ground state properties~\cite{kinoshita2004, paredes2004},  quantum dynamics~\cite{kinoshita2006,hofferberth2007,palzer2009, fabbri2012,kuhnert2013,ronzheimer2013}, and quantum correlation~\cite{cazalilla2011, guan2013}. Interestingly, a natural extension of the LL model is to include the long-range dipole-dipole interaction (DDI) which presents for atoms with large magnetic dipole moment~\cite{Chomaz2023}. Although DDI makes the 1D system non-integrable, it enriches the quantum phases of the system. Theoretically, 1D dipolar gases have been extensively studied in using various theoretical approaches. Arkhipov {\it et al.} studied the ground-state properties of an 1D system of dipoles by means of a quantum Monte Carlo method~\cite{arkhipov2005}, it was shown that the system supports a crossover from a liquidlike to a solidlike state. Deuretzbacher {\it et al.} study the ground state of few bosons with repulsive dipole-dipole interaction in a quasi-one-dimensional harmonic trap by means of the exact diagonalization method~\cite{deuretzbacher2010ground}, in which it is predicted that there are three regimes roughly corresponding to these found in Ref.~\cite{rincon2015}. Citro {\it et al.} confirmed the Luttinger-liquid behavior of the 1D dipolar gases using reptation quantum Monte Carlo method~\cite{citro2007evidence}. De Palo {\it et al.} calculated the ground-state energy of quasi-1D dipolar gases using a variational approximation based on the Bethe ansatz ground state wave function of the Lieb-Liniger model~\cite{Palo2020}. More importantly, there have been experimental developments involving highly magnetic dysprosium atoms in quasi-one dimension~\cite{kao2021,li2023}.  For instance, topological pumping of the 1D dysprosium gases into strongly correlated nonthermal excited states are demonstrated~\cite{kao2021} and the rapidity and momentum distributions of quasi-1D dysprosium gases were experimentally measured~\cite{li2023}.

As an efficient representation for quantum states in low-dimensional systems, the continuous matrix product state (CMPS) method offers a unique opportunity for studying the ground-state and dynamic properties of these systems~\cite{verstraete2010, haegeman2010,haegeman2013, draxler2013, quijandria2014,eichler2015, ganahl2017}. Notably, CMPS has been successfully applied in the study of one-dimensional Bose gases~\cite{rincon2015,  lukin2022}, Fermi gases~\cite{chung2015}, and Bose-Fermi mixtures~\cite{peacock2022}.  In particular, Rinc\'on {\it et al.} study the ground-state phases of the Lieb-Liniger model with exponentially decaying interactions~\cite{rincon2015}. Other than the superfluid phase found in the usual LL model, the exponentially decaying interaction supports the strongly correlated super–TG and quasicrystal phases. Nevertheless, its potential within the context of quasi-1D dipolar systems remains largely unexplored.

In the present work, we investigate the ground-state properties of the quasi-1D dipolar gases with repulsive contact and dipolar interactions using the CMPS.  The interplay between short-range s-wave interactions and long-range DDI in atomic gases has opened the door to exploring a wide variety of phenomena.
Following Ref.~\cite{rincon2015}, we compute the first- and second-order correlation functions and use them to identify quantum phases of the gases by following the Luttinger liquid theory. We show that there exists three distinct quantum phases, i.e., the superfluid, super-TG, and quasicrystal phases. In general, the strong repulsive contact interaction produces a TG-type fermionization; while the large dipolar inteaction leads the system into a strongly correlated super–TG and quasicrystal phases. We also map out the phase diagram on the parameter plane consisting of the repulsive contact and dipolar interaction strengths. It turns out that the transitions between adjacent phases are of the crossover type. Finally, we discuss the experimental feasibility by estimating the dipolar interaction strength in both atomic and molecular gases. We find that although the predicted dipolar effects can hardly be detected in dipolar atomic gas, it is highly possible to observe the strongly correlated phases in ultracold gases of polar molecules.

This paper is organized as follows. In Sec.~\ref{model}, we introduce our model. In Sec.~\ref{resu}, we present the numerical results which include the phase diagram, Luttinger parameter, correlation functions, structure factor,  and momentum distribution. Finally, we conclude in Sec.~\ref{concl}.



\section{Quasi-1D dipolar Bose gases}\label{model}
We consider a gas of dipolar bosons confined in a transverse harmonic potential, i.e., $$U_\perp(y,z)=\frac{1}{2}m\omega_\perp^2(y^2+z^2),$$
where $m$ is the mass of the atom and $\omega_\perp$ is the frequency of the trap. To form the quasi-1D geometry, we assume that $\omega_\perp$ is sufficiently large such that the transverse motion of the atoms is frozen to the ground state of the harmonic potential. After integrating out the transverse degrees of freedom, we obtain the quasi-1D Hamiltonian
\begin{align}
\hat{H}&=\hat{H}_{\mathrm{kin}}+\hat{H}_{\mathrm{int}},
\end{align}
where the kinetic and interaction terms are, respectively,
\begin{align}
\hat{H}_{\mathrm{kin}}& =\int_{-\infty}^\infty dx\hat{\psi}^{\dagger }(x)\left( -\frac{\hbar
^{2}}{2m}\frac{d^{2}}{dx^{2}}\right) \hat{\psi}(x),  \\
\hat{H}_{\mathrm{int}}& =\frac{1}{2}\int_{-\infty}^\infty dxdx^{\prime }\hat{\psi}^{\dagger
}(x)\hat{\psi}^{\dagger }(x^{\prime })V_{\mathrm{int}}(x-x^{\prime })\hat{%
\psi}(x^{\prime })\hat{\psi}(x).
\end{align}
Here $\hat{\psi}(x)$ is the field operator and
\begin{align}
V_{\mathrm{int}}(x)=g_0^{(\mathrm{1D})}\delta(x)+V_{\mathrm{dd}}^{(\mathrm{1D})}(x)
\end{align}
is the interatomic potential which consists of the contact interaction and DDI. More specifically,  $g_0^{(\mathrm{1D})}=-2\hbar^2/(m a_{\mathrm{1D}})$ is the strength of the contact interaction with $a_{\mathrm{1D}}=-\ell_\perp\left(\ell_{\perp}/a_{\mathrm{3D}}-\mathcal{C}/\sqrt{2}\right)$. Here $\ell_{\perp}=\sqrt{\hbar/(m\omega_\perp)}$ is the width of the transverse harmonic oscillator, $\mathcal{C}=1.4603$ is a constant, and $a_{\mathrm{3D}}$ is the $s$-wave scattering length in three dimensions. Furthermore, the quasi-1D dipolar interaction contains a long-range and a short-range parts~\cite{deuretzbacher2010ground,sinha2007cold, deuretzbacher2013}, i.e.,
\begin{align}
V_{\mathrm{dd}}^{(\mathrm{1D})}(x)=\frac{g_{L}}{\ell_\perp^3}\left[v_{L}(x/\ell_\perp)-%
\frac{8}{3}\delta(x/\ell_\perp)\right],
\end{align}
where the DDI strength is $g_{L}=c_{\rm dd}(1-3\cos^2\theta)/4$ with $\theta$ being the angle between the dipole moment and the $x$ direction and $c_{\rm dd}=\mu_0d^2/(4\pi)$ or $d^2/(4\pi\varepsilon_0)$ for, respectively, the magnetic or electric dipoles. Here $d$ is the dipole moment and $\mu_0$ ($\varepsilon_0$) is the vacuum permeability (permittivity). Finally, the long-range part of the quasi-1D DDI potential is
\begin{align}
v_{L}(u)=-2|u|+ \sqrt{2\pi}(1+u^2){e}^{u^2/2}\mathrm{erfc}\left(\frac{| u|}{%
\sqrt{2}}\right),  \label{intlong}
\end{align}
where $\mathrm{erfc}(\cdot)$ is the complementary error function. We point out that $v_L(u)$ is a monotonically decreasing function of $u$ and $v_L(u)$ is finite as $|u|\rightarrow0$.

It is convenient to repack the interaction potential into the sum of short- and long-ranged parts, i.e.,
\begin{align}
V_{\mathrm{int}}(x)=g_{S}\delta(x)+\frac{g_{L}}{\ell_{\perp}^3}%
v_{L}(x/\ell_{\perp}),  \label{totintpot}
\end{align}
where
\begin{align}
g_{S}&=g_{0}^{(\mathrm{1D})}-\frac{8g_{L}}{3\ell_\perp^2}=-\hbar\omega_\perp%
\left[\frac{\ell_\perp^2}{a_{\mathrm{1D}}}+a_{\mathrm{dd}}(\theta)\right]
\end{align}
is the strength of the short-range interaction that combines the contributions from the collisional and dipolar interactions. Because $g_{S}$ and $g_{L}$ are highly tunable quantities, quasi-1D dipolar gases can represent a broad range of models. For instance, the system represents the Lieb-Liniger model if $g_{L}=0$. In particular, in the limit of $g_{S}\to\infty$, it reduces to the Tonks-Girardeau (TG) gas, for which the bosonic particles effectively behave as free fermions.

\subsection{Scaling analyses}
To reveal some fundamental properties of our system, let us first recall the CMPS method for bosons introduce in by Verstraete and Cirac~\cite{verstraete2010}. A CMPS wave function for an 1D bosonic system of length $L$ with periodic boundary condition takes the form~%
\cite{verstraete2010}
\begin{align}
|\Psi \rangle =\mathrm{tr}\left[ \mathcal{P}e^{\int_{0}^{L}[Q(x)\otimes \hat{I}
+R(x)\otimes \hat{\psi}^{\dagger }(x)]\mathrm{d}x}\right] |0\rangle ,\label{def_CMPS}
\end{align}
where $\hat{I}$ is the unit operator, $Q(x)$ and $R(x)$ are ${B\times B}$ matrices in the $B$-dimensional auxiliary space introduced as the variational parameters of the
system, $\mathrm{tr}_{\text{aux}}(\cdot )$ represents the trace over the
auxiliary system, $\mathcal{P}$ denotes the path-ordering, and $|0\rangle $
is the vacuum state. Here, we concentrate on systems with translation invariance, therefore, $\{Q,R\}$ are independent of the position $x$.

The variational parameters $\{R,Q\}$ can be determined by minimizing the total energy
\begin{align}
E(n,g_S,g_L)&=E_{\mathrm{kin}}+E_{\mathrm{int}}\label{totale}
\end{align}
where $n= \langle\Psi|\hat\psi^{\dagger}(x) \hat\psi(x)|\Psi\rangle/\langle\Psi |\Psi \rangle$ is the density of the gas, $E_{\mathrm{kin}}=\langle \Psi |\hat{H}_{\mathrm{kin}}|\Psi \rangle /\langle\Psi |\Psi \rangle $ is the kinetic energy and $E_{\mathrm{int}}=\langle\Psi |\hat{H}_{\mathrm{int}}|\Psi \rangle /\langle \Psi |\Psi \rangle $ is the interaction energy. It should be note that, to write out Eq.~\eqref{totale}, we have made use of the fact that the total energy is a function of $n$, $g_S$, and $g_L$.

Following Ref.~\cite{verstraete2010}, we note that there exists a scaling relation between systems
with different number densities. To show this, it is convenient to decompose the interaction energy into the short- and long-range parts and re-express it as $E_{\mathrm{int}}=g_S\mathcal{E}_{\mathrm{sr}}+g_L\mathcal{E}_{\mathrm{lr}}$. Then, as shown in Ref.~\cite{verstraete2010}, there exists a scaling transformation
\begin{align}
Q\rightarrow \lambda Q\mbox{ and }R\rightarrow \sqrt{\lambda}R,
\label{scaling}
\end{align}
under which the physical quantities transform according to $n\rightarrow\lambda n$, $E_{\mathrm{kin}}\rightarrow \lambda^3E_{\mathrm{kin}}$, and $\mathcal{E}_{\mathrm{sr}}\rightarrow\lambda^2E_{\mathrm{sr}}$. As a result, in the absence of the long-range interaction, the total energy for a system with number density $n$ relates to that with unit density through the relation~\cite{verstraete2010}
\begin{align}
E(n,g_S,g_L=0)=n^3E(n=1,g_S/n,g_L=0).\label{relx3}
\end{align}
This relation greatly simplify the numerically calculation as we always fix number density at $n=1$.

In the presence of DDI, because the long-range interaction $v_L(x)$ in Eq.~\eqref{intlong} does not lead to a simple transformation relation for $\mathcal{E}_{\mathrm{lr}}$ under the scaling transformation~\eqref{scaling}, one cannot relate the total energies for systems with different number densities. Nevertheless, we can still find a scaling relation in the limit $\ell_\perp\rightarrow0$ under which we have $\ell_\perp^{-3}v_L(x/\ell_\perp)\rightarrow x^{-3}$. Then under the scaling transformation~\eqref{scaling}, we have $\mathcal{E}_{\mathrm{lr}}\rightarrow\lambda^4\mathcal{E}_{\mathrm{lr}}$, which leads to
\begin{align}
E(n,g_S,g_L)=n^3E(n=1,g_S/n,ng_L)\mbox{ for $\ell_{\perp}\rightarrow0$}.
\notag
\end{align}
In Sec.~\ref{resu}, we shall verify this scaling relation through numerical calculations.

\section{Numerical results}\label{resu}
In this section, we explore the ground-state properties of the quasi-1D dipolar gases with a given density $n$. To this end, we first introduce a chemical potential $\mu$ and then numerically minimize the free energy $E-\mu n$.

For convenience, we introduce a set dimensionless units: $\hbar\omega_\perp$ for energy, $\ell_\perp$ for length, and $\ell_\perp^{-1/2}$ for field operator. As a result, the Hamiltonian, in the dimensionless form, takes the familiar form:
\begin{align}
\bar H&=\int_0^\infty d\bar x \frac{d\bar{\psi}^{\dagger }(\bar{x})}{d\bar{x}}\frac{d\bar{\psi}(\bar x)}{d\bar{x}}\nonumber\\
&\quad+\int_0^\infty d\bar xd\bar x^{\prime }\bar{\psi}^{\dagger
}(\bar x)\bar{\psi}^{\dagger }(\bar x^{\prime })\bar V_{\mathrm{int}}(\bar x-\bar x^{\prime })\bar{%
\psi}(\bar x^{\prime })\bar{\psi}(\bar x),
\end{align}
where
\begin{align}
\bar V_{\mathrm{int}}(\bar x)=C\delta(\bar x)+Dv_L(\bar x)
\end{align}
with $C=g_S/(\hbar\omega_\perp\ell_\perp)$ and $D=g_L/(\hbar\omega_\perp\ell_\perp^3)$ being the dimensionless strengths for the contact and dipolar interactions, respectively. For the short-hand notation, we shall drop the `bar' over the dimensionless quantities, which should not cause any ambiguity.

To characterize the ground-state properties of the gases, we compute the first- and second-order correlation functions, i.e.,
\begin{align}\label{def_g2}
g^{(1)}(x)&=\frac{1}{n}\langle\hat\psi^{\dagger}(x) \hat\psi(0)\rangle
\end{align}
and
\begin{align}\label{def_g2}
g^{(2)}(x)&=\frac{1}{n^2}\langle \hat\psi^{\dagger}(0) \hat\psi^{\dagger}(x) \hat\psi(x)\hat\psi(0)\rangle,
\end{align}
respectively. Clearly, $g^{(1)}(x)$ characterizes the superfluid correlation and $g^{(2)}(x)$ represents the pair correlation. According to the Luttinger liquid theory, the asymptotic expressions for these correlations at long distances, $nx\gg 1$, are give by~\cite{rincon2015, giamarchi2003, cazalilla2004}
\begin{align}
g^{(1)}(n x) &\approx \frac{1}{(n x)^{1/(2K)}}\left[B_0 + B_1\frac{\cos(2\pi n x)}{(n x)^{2K}}\right],\label{g1ana}\\
g^{(2)}(n x) &\approx 1-\frac{K}{2\pi^2}\frac{1}{(n x)^2}+ A_1 \frac{\cos(2\pi n x)}{(n x)^{2K}},\label{g2ana}
\end{align}
where $K$, $A_1$, $B_0$, and $B_1$ are coefficients that can be fitted from the numerical results. Physically, the values of these coefficients depend on the microscopic parameters in Hamiltonian. In particular, $K$ is the Luttinger parameter employed to characterize the state of the gas. Here we will follow the convention of calling the ground state of model (1) superfluid when $g^{(1)}(x)$ decays slower than $g^{(2)}(x)$. This will be satisfied if $K > 1/2$. Similarly, we will say that the ground state has charge order if $K < 1/2$, i.e., whenever $g^{(2)}(x)$ decays slower than $g^{(1)}(x)$. This convention stems from the fact that in one dimension there is no breaking of continuous symmetries; hence algebraically decaying correlations are the closest behavior to long-range order. Since the Tonks-Girardeau (TG) gas corresponds to $K=1$, we shall follow the convention in Ref.~\cite{rincon2015} and refer the $1/2>K>1$ as the super-TG gas.

As an experimentally accessible quantity using Bragg spectroscopy, the static structure factor
\begin{align}  \label{static structure factor}
S(k)=1+n\int^{\infty}_{-\infty} dx {e}^{-i kx}\left[g^{(2)}(x)-1\right]
\end{align}
is also of great importance. In particular, it provides information about the spatial arrangement of particles in many-body systems, which can be used to determine whether the system has a crystal structure. Finally, we shall also  compute the momentum distribution
\begin{align}
p(k)&=\frac{n}{2\pi\hbar}\int_{-\infty}^\infty dxe^{-ikx}g^{(1)}(x),
\end{align}
which was experimentally measured for quasi-1D dipolar gases~\cite{li2023}.

Below we shall first check the validity of the CMPS calculations and then present our results on the ground-state properties of the quasi-1D dipolar gases. We also point out that, for the results presented in this work, the gas density is fixed at $n=1$ unless otherwise stated. This is equivalent to choose $n^{-1}$ as the length unit.

\subsection{Validity checks}

\begin{figure}[ptb]
\centering
\includegraphics[width=0.85\columnwidth]{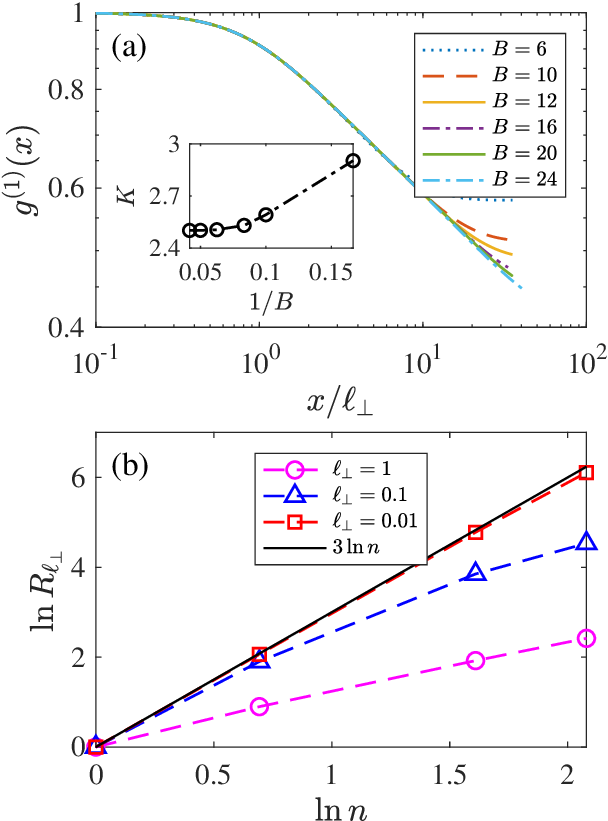}
\caption{(color online). Validity checks for the numerical calculations. (a) Superfluid correlation functions calculated using different bond dimensions. Other parameters are $n=1$, $C=2$, and $D=0.1$. (b) Ratio $R_{\ell_\perp}$ as a function of density $n$ for $C=0$, $D=10$, and various $\ell_{\perp}$'s. For reference we have also plotted the function $n^3$.}
\label{validity}
\end{figure}

As the preliminary checks for our CMPS calculations, we first solve the Lieb-Liniger model ($D=0$) by following Refs.~\cite{verstraete2010,rincon2015}. Unsurprisingly, our calculations can reproduce those presented in these studies. For the next step, we perform the validity checks in the presence of the long-range dipolar interaction. Specifically, we examine the convergence of the CMPS wave function by increasing the bond dimension $B$. In Fig.~\ref{validity}(a), we compare the first-order correlation functions computed using different bond dimensions for the set of parameters $n=1$, $C=2$, and $D=0.1$. As can be seen, the correlation functions converge up to $x\sim30$ for $B=24$. And the size of the power-law region in $g^{(1)}(x)$ also increases with $B$. Moreover, it is found that all physical quantities converges with $B$. As an example, we plot the Luttinger parameter $K$ as a function of $B$ in the inset of Fig.~\ref{validity}(a). Clearly, $K$ exhibits a systematic convergence as $B$ increases. Following Rinc\'on {\it et al}.~\cite{rincon2015}, we assume that $K(B)$ is quadratic in $1/B$. As a result, the relative error for $B=24$ is as low as $2.2\%$. In numerical calculations, it is found that the calculation is extremely time consuming for $B=24$, we shall use $B=16$ for all results presented in this work. We point out that the relative error with $B=16$ is still below $3\%$.

In addition to checking the convergence of the CMPS solution for quasi-1D dipolar gases, the correctness of these solutions should also be examined. Unlike the integrable Lieb-Liniger model which can be exactly solved using Bethe ansatz, the direct comparison to exact solution for the non-integrable dipolar gases is inapplicable. Here we present an indirect check by verifying the scaling relation Eq.~\eqref{relx3}. As an example, we plot, in Fig.~\ref{validity}(b), the ratio
\begin{align}
R_{\ell_\perp}\equiv \frac{E(n,C,D)}{E(n=1,n^{-1}C,nD)}
\end{align}
as a function of $n$ for various $\ell_\perp$'s with $C=0$ and $D=10$. As can be seen, $R_{\ell_\perp}$ approaches to $n^3$ as $\ell_\perp\rightarrow0$, which verifies the scaling relation Eq.~\eqref{relx3} and indirectly proves the validity of the CMPS calculations

\subsection{Luttinger parameter}
\begin{figure}[tbp]
\centering
\includegraphics[width=0.85\columnwidth]{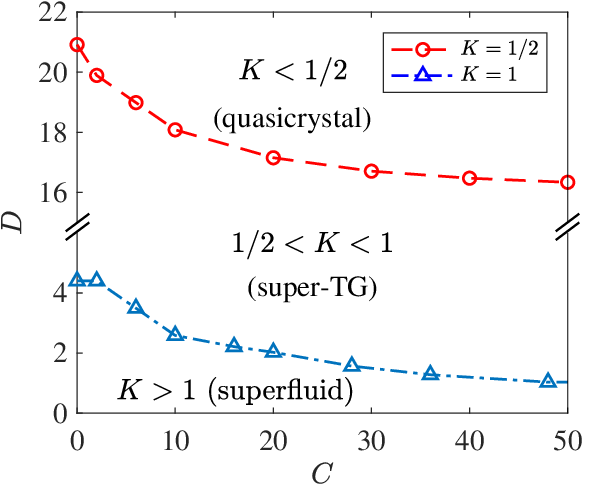}
\caption{Quantum phases on the $C$-$D$ plane characterized by the Luttering parameter $K$. The markers $\ocircle$ and $\triangle$ denote the boundaries determined by $K=1/2$ and $1$, respectively. The regions with $K<1/2$ and $K>1$ belong to the quasicrystal and superfluid phases, respectively; while that with $1/2<K<1$ is the crossover region. The Tonks-Girardeau gas corresponds to $K = 1$.}
\label{phase}
\end{figure}

Figure~\ref{phase} summarizes the main results of this work by mapping out the quantum states of the quasi-1D dipolar gases on the $C$-$D$ plane. As can be seen, the parameter plane is divided into three regions by the lines defined by the equations $K(C,D)=1/2$ and $1$. According to the criterion proposed in Ref.~\cite{rincon2015}, the regions satisfying $K<1/2$ and $K>1$ belong, respectively, to the quasicrystal and superfluid states, and the intermediate region with $1/2<K<1$ is the super-TG state. To facilitate further discussion, we introduce two critical dipolar interaction strengths, $D_{1/2}^*(C)$ and $D_{1}^*(C)$, corresponding to the lines denoted, in Fig.~~\ref{phase}, by $\ocircle$ and $\triangle$, respectively. Apparently, both $D_{1/2}^*(C)$ and $D_{1}^*(C)$ are monotonically decreasing functions of $C$. For $C=0$, the critical dipolar strength for the super-TG phase is $D_1^*(0)\approx4.2$. The presence of the contact interaction then bring $D_1^*$ down to zero at the large $C$ limit, i.e., $D_1^*(\infty)=0$, which states that the TG regime can only be achieved in the limit $C\rightarrow\infty$ without DDI. Moreover, the quasicrystal state only appears when $D$ is sufficiently large. Although the contact interaction alone does not leads to quasicrystal, its presence is favorable for quasicrystal state such that the minimium dipolar interaction strength is $D_{1/2}^*(\infty)\approx 16.1$. Then lowering $C$ leads to a larger critical dipolar strength for quasicrystal such that $D_{1/2}^*(0)\approx21$.

\begin{figure}[tbp]
\centering
\includegraphics[width=0.85\columnwidth]{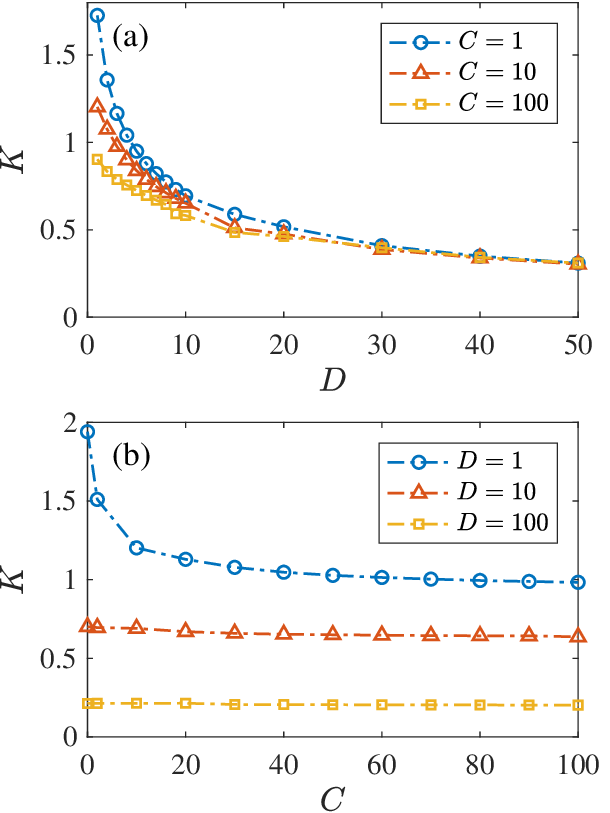}
\caption{(color online). Luttinger parameters. (a) $K$ versus $D$ for various $C$'s. (b) $K$ versus $C$ for various $D$'s.}\label{luttpara}
\end{figure}

To gain more insight into the behavior of the Luttinger parameter, we plot the $D$ dependences of $K$ in Fig.~\ref{luttpara}(a). As can be seen, the value of the Luttinger parameter may cover all three phases by varying $D$. In principle, the Luttinger parameter ranges over the whole positive real axis, in analogy to that of the extended Lieb-Liniger (ELL) model~\cite{rincon2015}. More specifically, for a given $C$, $K$ is a monotonically decreasing function of $D$. Even though $K$ decreases more rapidly when $C$ is smaller, all $K$ converge to the same value ($< 1/2$) as $D\rightarrow\infty$. This suggests that the properties of the quasicrystal phase in the large $D$ limit is solely determined by the long-range interaction. Furthermore, we present, in Fig.~\ref{luttpara}(b), the Luttinger parameter $K$ as a function of $C$ for different $D$'s. As can be seen, for a given $D$, $K$ is always a decreasing function of $C$ which, at large $C$ limit, converges to a value ($\leq 1$) solely determined by $D$. In addition, the larger the $D$ is, the smaller the slope of the decreasing function $K(C)$ becomes. And $K(C)$ is roughly independent of $C$ when $D=100$.

\subsection{Correlation functions}
\begin{figure}[tbp]
\centering
\includegraphics[width=0.85\columnwidth]{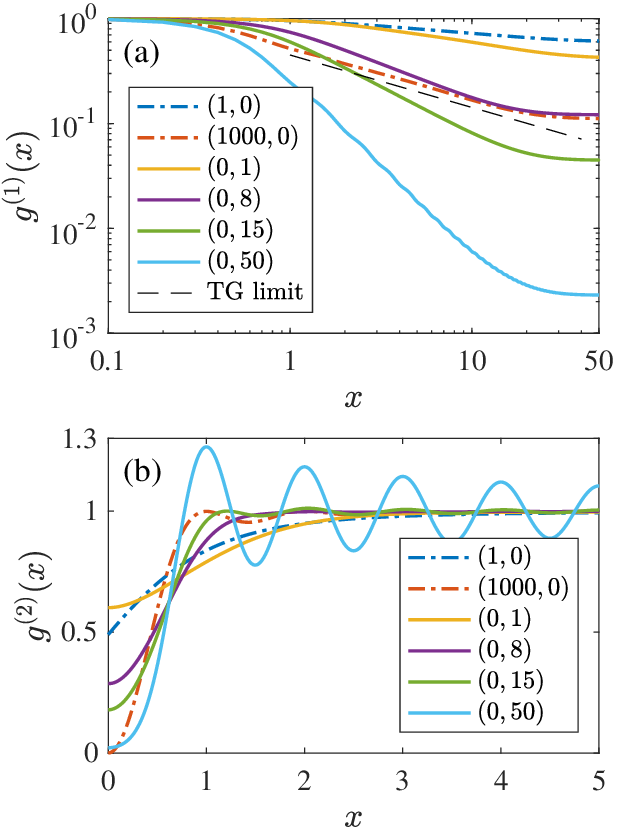}
\caption{(color online). Superfluid (a) and pair (b) correlation functions for various sets of interaction parameters $(C,D)$. The straight dashed line with a slope $1/2$ in (a) denotes the TG limit.}
\label{g1g2}
\end{figure}

To further explore the properties of the quasi-1D dipolar gases, let us examine the correlation functions of the quantum states. To this end, we plot the distance dependences of the superfluid and pair correlations in Fig.~\ref{g1g2}(a) and (b), respectively, for various sets of interaction strengths $(C,D)$. For a better comparison, the parameter sets are chosen to represent the LL models (dash-dotted lines) and the pure dipolar models (solid lines).

For comparison, let us briefly recall the correlations of the LL model~\cite{astrakharchik2006}. For the weak coupling case with $C=1$, $g^{(1)}(x)$ clearly demonstrates superfluid order. As $C$ increases, the superfluid correlation is gradually suppressed and, in the strong coupling case ($C=10^3$), the superfluid correlation exhibits the behavior of the TG limit which corresponds a spinless fermionic system. As to the pair correlation of the LL model, $g^{(2)}(x)$ of $C=1$ monotonically increases from its minimum value $g^{(2)}(0)$ to the asymptotic value at large distance. While for large $C$, the Friedal oscillation is developed on $g^{(2)}(x)$, which is the characteristic of the Tonks-Girardeau regime. In fact, the pair correlation function of $C=10^3$ is visually indistinguishable from that of the TG gas, i.e., $g^{(2)}(x)=1-\sin^2\pi nx/(\pi nx)^2$, indicating that $C=10^3$ is sufficient to represent the strongly coupled LL model.

As to the pure dipolar model, the interaction strengths adopted in Fig.~\ref{g1g2} cover all the regions of interest. Let us first examine the superfluid correlation. For the weak coupling case ($D=1$), $g^{(1)}(x)$ closely resembles those of the the LL case with $K>1$ such that the system is also in the superfluid state. For $D=8$, the superfluidity is suppressed although the gas is still in the superfluid state. In addition, because the Luttinger parameter satisfies $1/2<K<1$, the bosonic system is described by strongly interacting repulsive spinless fermions, i.e., a suppressed superfluid state or the super–TG regime. Further increasing $D$ leads to an almost complete suppression of superfluid correlations such that the Luttinger parameter reduces to $K<1/2$.

For the pair correlation function, $g^{(2)}(x)$, of the pure dipolar model, its long-distance behavior for $D=1$ is very similar to that of the LL model in the weak-coupling regime. Surprisingly, we do not observe any Friedal oscillation for $D=8$ even though the Luttinger parameter is smaller than unit. This suggests that the short-range repulsion induced by the long-range potential $v_L(x)$ is not as strong as the contact interaction since $v_L(x)$ is finite as $|x|\rightarrow0$. For $D=15$, although Friedal oscillation appears on $g^{(2)}(x)$, the pair correlation function for pure dipolar gas differs significantly from that of the spinless fermions (i.e., the $C=10^3$ case) at short distance. However, for $x\apprge3$, these two becomes nearly indistinguishable from each other. Further increasing $D$, the oscillation on $g^{(2)}(x)$ becomes more pronounced. Of particular importance, the pair correlations decays slower than the superfluid correlation, indicating that there exists a definite wave vector modulating the density fluctuations. The appearance of this wave vector in $g^{(2)}(x)$ signals the establishment of charge order, i.e., a quasicrystal state.

\subsection{Structure factor and momentum distribution}

\begin{figure}[tbp]
\centering
\includegraphics[width=0.8\columnwidth]{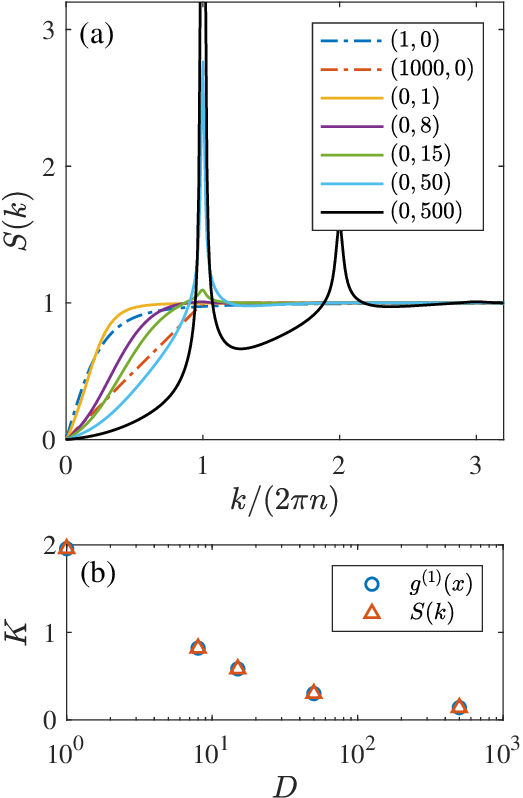}
\caption{(color online). (a) Static structure factor $S(k)$ for various sets of interaction parameters $(C,D)$. (b) Comparison of the Luttinger parameters fitted using $g^{(1)}(x)$ and $S(k)$ for pure dipolar model.}
\label{strufac}
\end{figure}

To gain more insight into the quasicrystal state, we examine the structure factor $S(k)$ which is experimentally measurable via the Bragg spectrometry. In Fig.~\ref{strufac}, we plot, in Fig.~\ref{strufac}(a), the structure factor $S(k)$ for various sets of parameters. For the LL model, the slope of $S(k)$ at the low-$k$ limit is $K/(2\pi n)$ which is proportional to the Luttinger parameter. In particular, the static structure factor of the TG gas takes an extremely simple form. Namely, $S(k)$ linearly increases until it reaches the asymptotic value $1$ at the wave vector $|k| = 2\pi n$. For the pure dipolar model, we may also use the low-$k$ behavior to fit the Luttinger parameter. As shown in Fig.~\ref{strufac}(b), the Luttinger parameter obtained by fitting the low-$k$ behavior of $S(k)$ is in good agreement with that obtained by fitting $g^{(1)}(x)$. For weak dipolar interaction ($D=1$), $S(k)$ resembles that of the LL model. While for $D=15$, a peak emerges at $k=2\pi n$, which confirms the charge order in the density fluctuations. As $D$ is further increased, not only the height of the peak increases, but addition peaks appears when $k/(2\pi n)$ is an integer.

\begin{figure}[tbp]
\centering
\includegraphics[width=0.85\columnwidth]{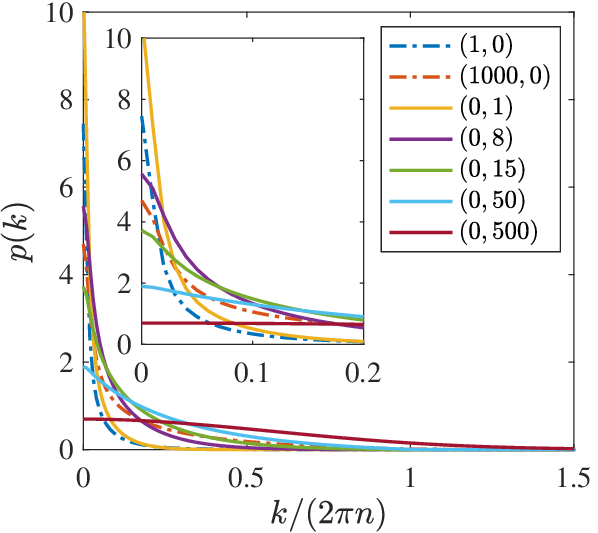}
\caption{(color online). Momentum distributions $p(k)$ for various sets of interaction parameters $(C,D)$. The inset is the zoom-in plot of the small $k$ regime.}
\label{moment}
\end{figure}

Finally, we examine the momentum distribution which was measured from the time-of-flight images for the a 1D dyspersium gas recently~\cite{li2023}. In Fig.~\ref{moment}, we present the momentum distributions for both LL model and pure dipolar gases. Ideally, as shown in Ref.~\cite{astrakharchik2006}, the momentum distribution diverges at $k\rightarrow0$ for the superfluid phase. However, the numerical results always indicate that $p(0)$ is finite. The underlying reason is that $g^{(1)}(x)$ decays very slowly such that it only vanishes until $x$ is so large that is inaccessible in numerical calculations. For dipolar gases, although $p(k)$ with small $D$ is similar to that of the LL model, the whole momentum distribution is significantly flattened as $D$ increases, which may facilitate the identifications of the dipolar effects in experiments.

\section{Conclusion and discussion}\label{concl}

In conclusion, we have presented a detailed study on the ground-state properties of the quasi-1D dipolar Bose gases by providing the phase diagram. In this system, the interplay of the contact and dipolar interaction gives rise to a variety of quantum phases characterized by the Luttinger parameter. Generally speaking, the strong repulsive contact interaction produces a TG-type fermionization; while the large dipolar inteaction leads the system into a strongly correlated super–TG and quasicrystal phases.

As to the experimental feasibility, let us estimate the dipolar interaction strengths for typical dipolar gases. Among the experimentally realized dipolar atomic gases, Dy atom possesses the largest magnetic dipole moment with $\mu_m=9.93\,\mu_B$, where $\mu_B$ is the Bohr magneton. Given the transverse trap frequency $\omega_\perp=2\pi\times 25\,\mathrm{kHz}$ as in the experiments~\cite{li2023}, the largest dimensionless dipolar interaction strength is $D\approx 0.1$, which is still too small for observing any dipolar effects in quasi-1D gases. However, we may also consider the quasi-1D gases of ultracold polar molecules~\cite{polmol1,polmol2,polmol3,polmol4,polmol5,polmol6,polmol7}. Provided that the typical electric dipole moment of these polar molecules is $3\,$Debye, the dimensionless dipolar strength can be as high as $D\approx 110$ with the same trap geometry as that of the atom's. Apparently, the strong dipolar interaction between polar molecules is sufficiently large for us to observe the quasicrystal phase.

\begin{acknowledgments}
We thank enlightening discussions with X.-W. Guan. This work was supported by National Key Research and Development Program of China (Grant No. 2021YFA0718304), by the NSFC (Grants No. 12135018, No. 12247118, and No. 12047503), and by the Youth Innovation Promotion Association.
\end{acknowledgments}

\end{document}